# COVID-19 Literature Mining and Retrieval using Text Mining Approaches

Sanku Satya Uday

*School of Engineering and Sciences*
*SRM University, Andhra Pradesh, India.*
*sanku_satya@srmap.edu.in*

Satti Thanuja Pavani

*School of Engineering and Sciences*
*SRM University, Andhra Pradesh, India.*
*satti_thanuja@srmap.edu.in*

T. Jaya Lakshmi

*School of Engineering and Sciences*
*SRM University, Andhra Pradesh, India.*
*jaya.phd.hcu@gmail.com*

Rohit Chivukula

*School of Computing and Engineering,*
*University of Huddersfield, United Kingdom.*
*rohit.chivukula.kingdom@gmail.com*

**Abstract**

The novel coronavirus disease (COVID-19) began in Wuhan, China, in late 2019 and to date has infected over 148M people worldwide, resulting in 3.12M deaths. On March 10, 2020, the World Health Organisation (WHO) declared it as a global pandemic. Many academicians and researchers started to publish papers describing the latest discoveries on covid-19. The large influx of publications made it hard for other researchers to go through a large amount of data and find the appropriate one that helps their research. So, the proposed model attempts to extract relavent titles from the large corpus of research publications which makes the job easy for the researchers. Allen Institute for AI released the CORD-19 dataset, which consists of 2,00,000 journal articles related to coronavirus-related research publications from PubMed's PMC, WHO (World Health Organization), bioRxiv, and medRxiv pre-prints. Along with this document corpus, they have also provided a topics dataset named topics-rnd3 consisting of a list of topics. Each topic has three types of representations like query, question, and narrative. These Datasets are made open for research, and also they released a TREC-COVID competition on Kaggle. Using these topics like queries, our goal is to find out the relevant documents in the CORD-19 dataset. In this research, relevant documents should be recognized for the posed topics in topics-rnd3 data set. The proposed model uses Natural Language Processing(NLP) techniques like Bag-of-Words, Average Word-2-Vec, Average BERT Base model and Tf-Idf weighted Word2Vec model to fabricate vectors for query, question, narrative, and combinations of them. Similarly, fabricate vectors for titles in the CORD-19 dataset. After fabricating vectors, cosine similarity is used for finding similarities between every two vectors. Cosine similarity helps us to find relevant documents for the given topic.

*Corresponding author
*Email address:* jaya.phd.hcu@gmail.com (T. Jaya Lakshmi)

*Keywords:* Covid-19, Bag-of-Words, Average Word-2-Vec, Average BERT Base, Cosine Similarity, Tf-Idf Weighted Word-2-Vec, Lemmatization

## 1. INTRODUCTION

As the pandemic has made its way to an extreme level, researchers worldwide are working to know the behavior of the virus and propose ways to impede the spread of the virus, therefore lakhs of research papers are generated. This massive generation of data makes it tough to find the relevant topic of interest for the researchers [1]. All this huge data was made available in the CORD-19 dataset by Allen Institute for AI [2]. Natural Language processing is a subfield of artificial intelligence concerned with computers and human language interactions, specifically how to program a computer to process and analyze large amounts of natural language data in a human-readable format [3]. This makes the computer capable of understanding the contents of documents and the context. Therefore, Natural Language Processing is applied to the dataset to analyze large amounts of research paper titles written in Natural Language and returns the relevant research papers for the topic of interest.

### *1.1. Problem Statement*

Given a large corpus of research documents related to COVID-19, each containing title and abstract and a list of topics at various granularity referred as query, question and narrative, the goal is to extract relevant research documents related to topic from the corpus.

The problem is challenging because there are multiple levels of granularity for both the document and the topic being searched for. Models for predicting the most relevant titles and abstracts for queries are proposed in this study. This allows researchers to acquire more accurate titles and abstracts for their queries.

## 2. Dataset Description

An information retrieval challenge with the name Trec-Covid has been organised by the Allen Institute for Artificial Intelligence, National Institute of Standards and Technology, National Library of Medicine, Oregon Health and Science University, and the University of Texas Health Science Center at Houston on the popular coding challenge platform Kaggle.com. Trec-Covid dispenses us with three data files named as metadata.csv (CORD-19 dataset), topics-rnd3.csv and qrels.csv.

Metadata.csv contains a unique identifier of each document called as *cord_uid*; along with the corresponding titles and abstracts. The abstract gives an extensive description of the content of the document. It also contains other features such as the *pubmed_id* of the authors and the *publish_time*. The metadata file consists of 1.3 lakh titles and abstracts.

The file topics-rnd3.csv contains topic-id, query, question, and narrative. The query is a short description of a topic, and the question is a slightly extensive description of a query, where as the narrative describes the question more in detail. A total of 40 topics have been made available by the challenge.

The file qrels.csv contains the relevant judgments for documents that have been evaluated in previous rounds.

The CORD-19 dataset consists of 103,000 documents in total. The dataset description is given in Table.1 and sample records in these three files are given in Tables 2,3 and 4.

Table 1: CORD-19 dataset description

| File | Description | #Records | #Features | Features |
|---|---|---|---|---|
| metadata.csv | Information about research articles published on Covid-19 | 103,000 | 19 | cord_id<br>title<br>abstract<br>pubmed_id<br>publish_time ... |
| topics-rnd3.csv | Queries to retrieve from documents at different levels of granularity. | 40 | 4 | topic_id<br>query<br>question<br>narrative |
| qrels.csv | Relevant judgments for documents that have been evaluated in previous rounds.<br>Three possible values of judgements:<br>0(non-relevant)<br>1(partially relevant) and<br>2(relevant). | 20,728 | 4 | topic_id<br>iteration<br>cord_id<br>judgement |

Table 2: Sample documents in CORD-19 Dataset (metadata.csv)

| Document No | cord_uid | Title | Abstract |
|---|---|---|---|
| 117 | rqkgrd2k | ebola virus come going | long text |
| 118 | q1q5801p | essentials pulmonology | long text |
| 119 | p6c57zzm | general mechanisms antiviral resistance | long text |

Table 3: Sample documents in qrls.csv of CORD-19 Dataset

| topic-id | iteration | cord-id | judgement |
|---|---|---|---|
| 1 | 0.5 | 010vptx3 | 2 |
| 1 | 1.0 | 02f0opkr | 1 |
| 1 | 1.0 | 04ftw7k9 | 0 |
| 1 | 1.0 | 05qglt1f | 0 |

Table 4: One sample document in topic.csv of CORD-19 Dataset

| Feature | value |
|---|---|
| **topic-id** | 1 |
| **query** | coronavirus origin |
| **question** | what is the origin of COVID-19 |
| **narrative** | seeking range of information about the SARS-Co... (long text) |

## 3. Related Literature

Machine learning is emerging as one of the hot topics in the industry as well as in academic institutions to smoothen the analytical tasks with huge amount of data being available daily. This data is mostly in the form of unstructured text documents [4]. There are algorithms in abundance for performing various analytical activities in order to gain actionable insights from this big data with the help of natural language processing (NLP) paradigm [4, 5].Data processing takes a long time, as demonstrated by the work of Long Ma et al., not only due to the sheer volume of information being processed, but also due to the variety and complexity of the data itself [6]. There are a variety of data mining and machine learning techniques being employed to address the problem of large data. These algorithms work effectively by choosing useful features.

*3.1. Structured representation of text data*

Bag of Words (BoW) is most popular model to transform unstructured form of text to structured [7]. BoW constructs fixed size numerical vector for each text document in the document corpus [8].

Word2Vec is a Google-proposed and funded algorithm that consists of two learning models: Continuous Bag of Words (CBOW) and Skip-gram. By feeding text data into one of the learning models, Word2Vec outputs word vectors that can be represented as a large piece of text or even the entire article [6].

BERT is a trained Transformer Encoder stack [9]. BERT receives a series of words as input, which continue to flow up the stack [9]. Each encoder performs self-attention, transfers the results through a feed-forward network, and then passes the information on to the next encoder. After reaching the last encoder the Bert base model releases the embedding for each word present in the sentence based on the context. The output will be 768 dimensional word embedding for each word. So for every sentence the number of word tensors depends upon the number of words present in that sentence.

Once the document vectors are constructed using the models described, the similarity can be computed using distance/similarity measures.

*3.2. Similarity measures*

Similarity is often assessed using a distance metric such as Euclidean distance or Cosine similarity. A basic metric for geometrical problems is Euclidean distance. It is the distance between two points and can be conveniently calculated in two or three dimensions. In clustering problems, such as text clustering, Euclidean distance is the commonly used measure [10]. Euclidean distance computes square root of summation of squared distance between the corresponding vector features. The general procedure of computing is shown in Fig.3.2.

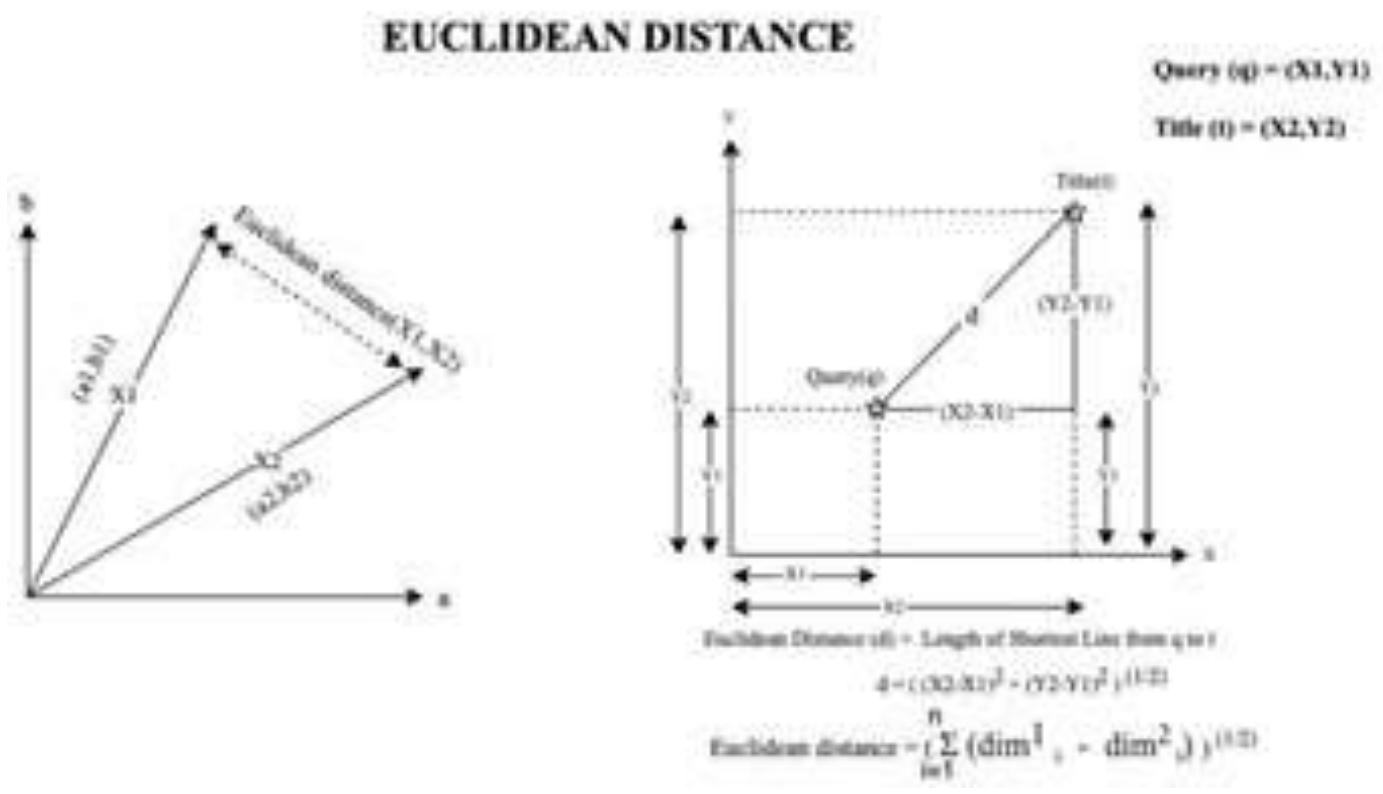

Figure 1: Euclidean Distance

Cosine similarity is another popular measure. Cosine similarity captures the correlation between the document vectors. This is expressed as the cosine of the angle formed by two vectors, or cosine similarity [11, 12]. Cosine similarity measures the angular distance between two vectors [13] which is shown in Fig.4.2.2.

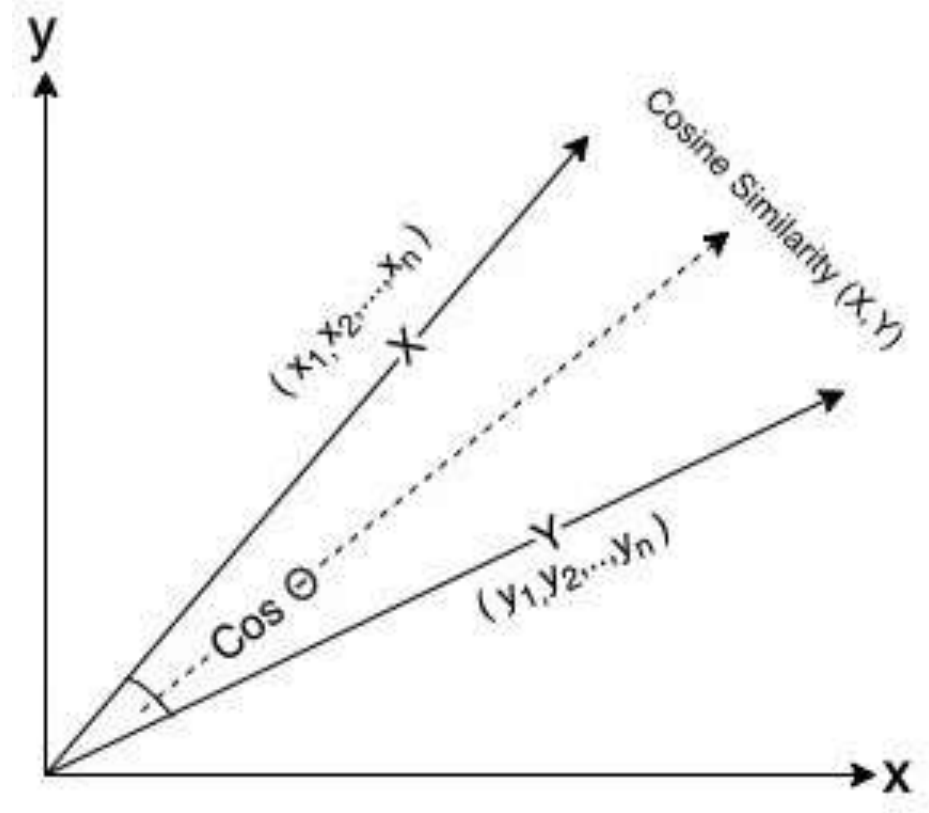

Figure 2: Cosine Similarity

Cosine similarity between a pair of vectors (X,Y) can be computed between two vectors X=($x_1$,$x_2$,. . . $x_n$) and Y=($y_1$,$y_2$,. . . $y_n$) using the equation 1.

$$
\begin{aligned}
&cos(x, y) = \frac{X.Y}{||X||_2||Y||_2} \\
&X.Y = x_1.x_2 \ldots x_n + y_1.y_2 \ldots y_n \\
&||X||_2 = x_1^2 + x_2^2 \ldots x_n^2 \\
&||Y||_2 = y_1^2 + y_2^2 \ldots y_n^2
\end{aligned}
\tag{1}
$$

Cosine similarity values range between 0 and 1. If the cosine similarity is zero, then the two documents being compared are dissimilar. The two documents are closely similar if cosine similarity is 1.

In this work, we use four natural language processing techniques to convert textual data to machine understandable language. The proposed approach is given in section 4.

## 4. COVID-19 Literature Mining and Retrieval

The main goal of this work is to output the titles of the documents that are most comparable to the posed queries. Firstly, the documents and queries are converted into vectors. Then similarity of each query is computed with each title. Once the similarity scores are available, the scores are sortedout to extract the top 'k' titles. The approach is shown in Fig.4.

**Algorithm 1** Relevant document extraction

Input:<br>
    'n' documents containing titles and abstracts<br>
    'm' Topics containing queries, questions and narratives.<br>
Output:<br>
    *Top_R Sim*: Top xx relevant titles for each query.<br>
    *Top_Q Sim*: Top xx relevant titles for each question<br>
    *Top_N Sim*: Top xx relevant titles for each narrative.<br>
**for** each $j$ in $1 \ldots m$ **do**<br>
    **for** each $i$ in $1 \ldots n$ **do**<br>
        Compute *R Sim* ← similarity($R_j$, $D_i$)<br>
        Compute *Q Sim* ← similarity($Q_j$, $D_i$)<br>
        Compute *N Sim* ← similarity($N_j$, $D_i$)<br>
    **end for**<br>
**end for**<br>
*Top_R Sim*: Top xx vectors from *R Sim*<br>
*Top_Q Sim*: Top xx vectors from *R Sim*<br>
*Top_N Sim*: Top xx vectors from *N Sim*<br>
Return *Top_R Sim*, *Top_Q Sim*, *Top_N Sim*

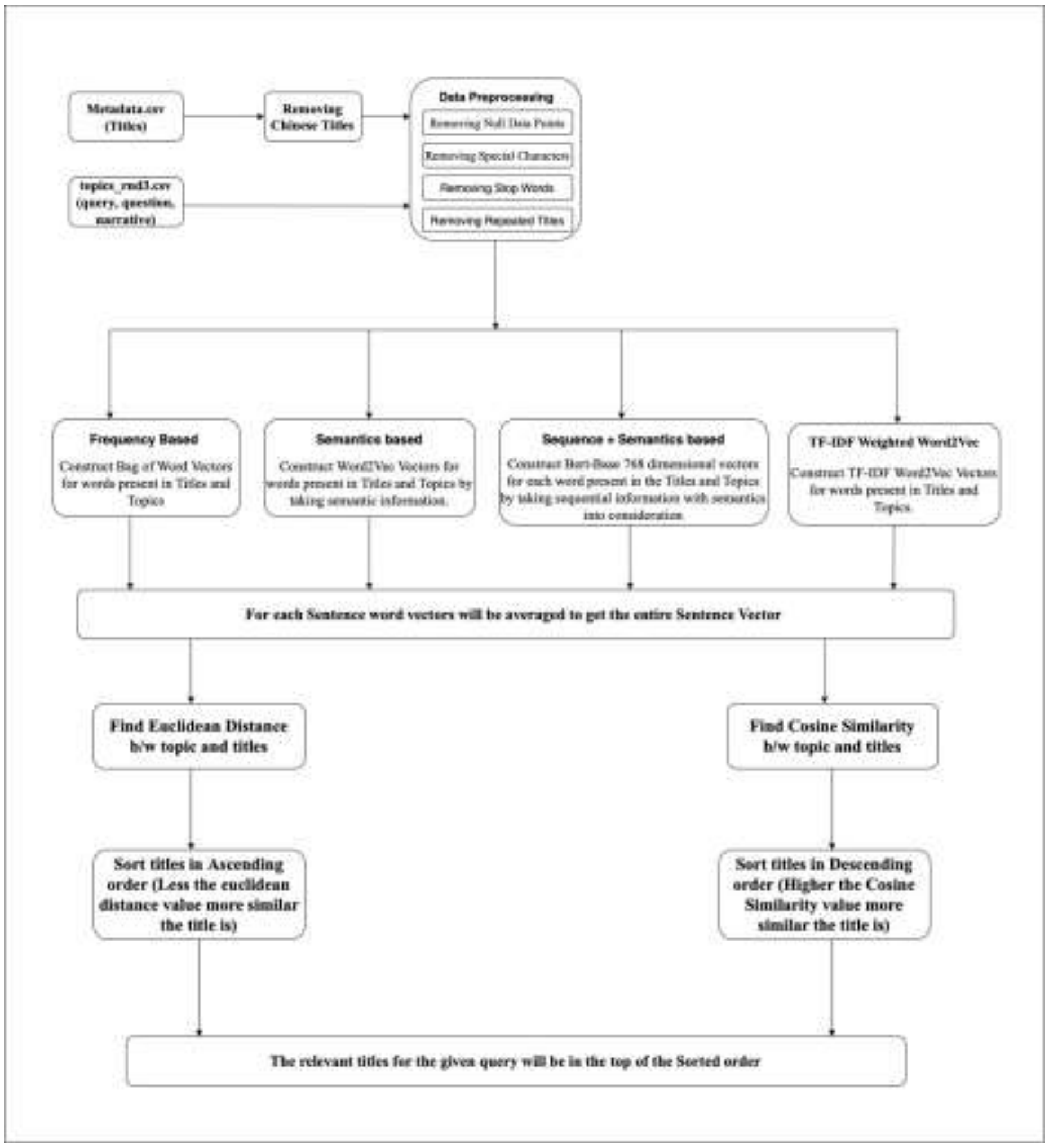

Figure 3: Proposed Approach

Steps in the approach are summarised as

- Preprocessing
- Apply NLP model
- Evaluate performance of the model

### *4.1. Preprocessing*

Titles in CORD-19 dataset need to be preprocessed for making it suitable for retrieval process.

Initially titles were taken from the CORD-19 dataset to observe the top 20 most perennial titles in the dataset. From this, it is noticed that there are 33 null titles. After removing these null titles, 1,28,459 titles were left in the dataset. Whenever the data points are removed from the dataset, the indices of those data points also get detached. So the index of the data frame must be reset after every removal. At present, all null data points were removed, and now special characters and stop-words must be removed that are nonsensical in the context of titles. Before removing these nonsensical words and symbols, transfigure every word in the title to lowercase to make succeeding evaluations or comparisons trouble-free. Previous to removing nonsensical words and characters, 6292 duplicated titles were noticed. However, there are 18064 repeated titles after preprocessing. This variability is because "[ Algemene leading ]" and "Algemene leading" are treated as two different titles since the first title contains special characters at the beginning and ending. After the removal of nonsensical words and characters from the titles, they are identical. Also, there will be no case-sensitive comparison trouble to get an accurate number of duplicated titles. Now titles were rearranged in ascending order. By this rearrangement, four null titles were identified. Possibly these four titles are made of stop words and special characters that make removal of the full title. In this way, arranging our data in ascending order helped us know the null titles and visualized the duplicate titles. After detaching duplicated titles and these four empty data points, ultimately getting 1,10,419 identical titles. Likewise, preprocess the topics-rnd3.csv

data by primarily transfiguring all the words into lowercase and then removing nonsensical stop-words and special characters present in queries, questions, and narratives. Once preprocessing completes, preprocessed data needs to be stored into a pickle file. Whenever preprocessed data is required, just put it back from the pickle file. There are a few Chinese titles in the given data set, but our processing is all on English language titles, hence Chinese titles will not fit into our processing. Therefore we have eliminated these using the procedure given in Section.4.1.1.

### *4.1.1. Non-English Titles*

While computing the bag-of-words, Chinese titles were captured while visualizing the heat maps of bag-of-word vectors in the CORD-19 dataset. To avoid these kinds of titles in the data set, the data set has been scanned using ASCII encoding.

Text elimination is commonly done using text comparison. But it is difficult to use the same process for non-English titles. Therefore utilizing ASCII encoding is useful in such scenarios.

An example of unreadable text in Chinese language taken from our experimentation is given in Fig.4.1.1

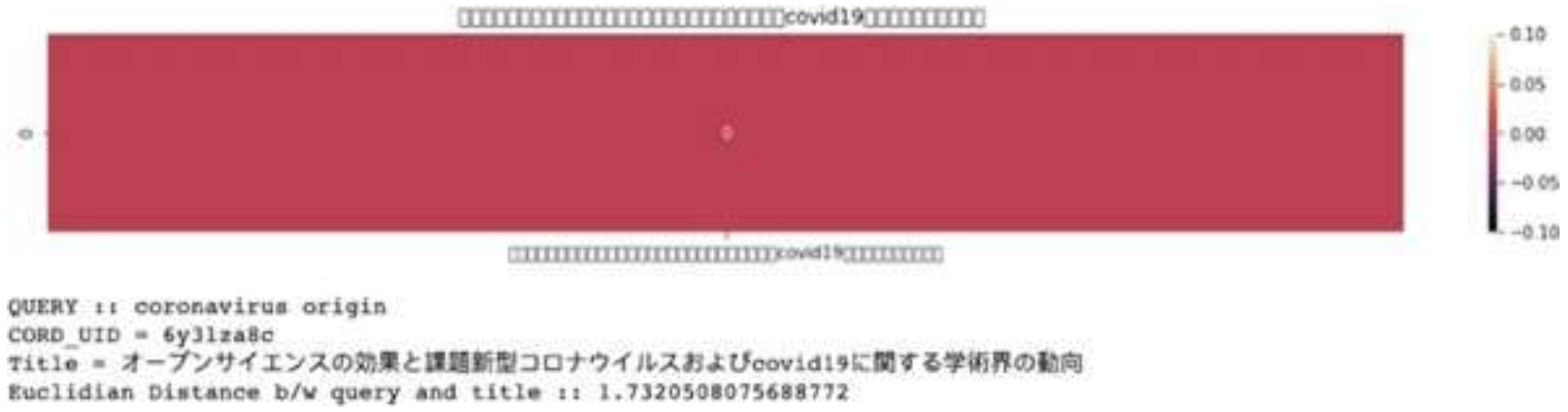

Figure 4: Chinese Titles

After preprocessing the given data through the methods mentioned in 4.1 and 4.1.1, Stemming and lemmatization techniques are used on these preprocessed titles and queries for reducing a word into its root form. The processes of stemming and lemmatization are briefed in the next two sub-sections.

### 4.1.2. Stemming

Stemming is the process of converting a word to its root form. Here porter stemmer is used for converting all the words in the titles into a root form [14]. Similarly, porter stemmer is used for queries, narratives, and questions. Then the bag of words is applied to fabricate the textual data into vectors. Finally cosine similarity is used for finding similarities between the titles and queries, for finding relevant titles for the given queries. The predictions made after stemming did not help us in achieving better accuracy.

### 4.1.3. Lemmatization

Lemmatization usually refers to doing things better by using a vocabulary and morphological analysis of words, typically aiming to remove inflectional endings only and return the base or dictionary form of the word [14]. Here Lemmatizer is used for converting all the words in the titles into a root form. Similarly, Lemmatizer is used for queries, narratives, and questions. Then bag-of-words is applied to fabricate the textual data into vectors. Finally cosine similarity is used for finding similarities between the titles and queries, for finding relevant titles for the given queries. Ultimately, applying Lemmatization on BoW vectors resulted in a greater accuracy that performed better than both stemming and similarities generated from non-lemmatized text. While in case of Word2Vec Lemmatization did not enhance the model performance. The lemmatized text helped only in frequency based model but not in semantic and sequence based models like Word2Vec, Bert-Base, Tfidf Weighted Word2Vec.

The preproccessed titles and queries are given to four different word embedding Natural Language Processing model's to fabricate the vectors for the preprocessed topics and titles. In this work, Bag of words, Word2Vec, Bert-Base model, Tf-Idf weighted Word2Vec are used to create the meaningful vectors for the given text data. Here Bag of Words and Word2Vec do not store the sequence information in the vectors whereas the Bert-Base model makes use of semantics with word sequence information to fabricate the vectors for the textual data. All these models output the vector's for every word present in the given text(title or topic). Since every text does not have the same number of words the model outputs a set of vectors for each title or topic but for computing the similarity between the topic and title, the dimensions of both the vectors should be the same. So, after creating the word vectors for each title or topic, all the word vectors corresponding to the given text are added up in a point wise fashion and then the average of every point in the vector is computed to get a unique one dimensional vector for the given text(title or topic). Finally, the topic vectors will be compared with title vectors with the help of Euclidean distance or Cosine Similarity to find out how title is closely relevant to the posed topic. The in depth working of the four Natural Language Processing(NLP) models is given in next four sections.

### *4.2. Bag of Words (BOW)*

The dataset contains titles of documents that are in the form of textual data. So transfigure the textual data into the form of vectors, which consist of frequency of words of that particular title. First of all, the preprocessed queries and titles are amalgamated into a single data frame. Promptly fabricate Bag of Word vectors for every title and query present in the single data frame with the help of CountVectorizer present in the sci-kit learn library. The procedure behind constructing a bag of words will take every title or query present in the data frame and takes the set of words present in the complete data frame of titles and queries. The set of words returned from the data frame is used as the bag of words vector features. Every title and query has the same dimensional Bag of words vector. Consider the title as $T_i$ and the words present in it as $W_j$ where j=0,1,...,n. The frequency of every $W_j$ present in the title $T_i$ is counted. Eventually, those frequency values are stored in the respective $W_j$ index present in the Bag of Word vector of title $T_i$. Similarly, frequency values were computed for every word in the title and stored in the respective title $T_i$'s vector. We use Euclidean distance as well as Cosine similarity for finding similarity between two vectors. The details of both these measures are given in Section 4.2.1 and Section 4.2.2.

#### 4.2.1. Euclidean Distance for Bag of Word Vectors

Initially, the Euclidean distance of query $Q_i$ with all the titles $T_j$ were computed with the help of the pairwise distances function. Using these euclidean distances, all indices of the titles having least euclidean distance are stored. Using these indices, the top titles with the least euclidean distance concerning the given query are returned.

Eventually, it is observed that the euclidean distances resulted in inadequate title suggestions for the given queries with an accuracy score of 6%. Promptly questions were taken instead of queries and then appended them to titles to perform a bag of words on both titles and questions. After computing bag- of-word vectors for questions and titles, it again resulted in inadequate title suggestions with an accuracy of 8%. Likewise, tried re-purposing input queries by taking the combinations of queries, questions, and narratives. Finally came to an epilogue that euclidean distance on a bag of word vectors leads to inadequate title suggestions with poor accuracy.

It is noticed that the euclidean distance resulted in inadequate title suggestions. In recent times cosine similarity became popular for finding better similarities of textual data.

#### 4.2.2. Cosine Similarity for Bag of Word Vectors

Cosine similarity measures the angular distance between the two BoW vectors. The higher the cosine similarity, the more similar the title is. So, titles will be sorted for a particular query based on the cosine similarities in descending order. This will help us to know the titles that are closely relevant to the query.

### 4.3. Average Word2vec

Word2vec is a technique for natural language processing. The word2vec algorithm uses a neural network model to learn word associations from a large corpus of text. Once trained, such a model can detect synonymous words or suggest additional words for a partial sentence [15]. Firstly word2vec model is trained by fabricating the word2vec's for every word in the given corpus (titles, queries, questions, narratives) with the help of the Word2vec model in the gensim.models library. Using this trained Word2vec model, for each word in a title 100 dimensional Word2vec's were fabricated. As Word2vec's for only words present in the titles or queries were computed. So now, Word2vec for the entire query or title needs to be fabricated. To fabricate the Word2vec for the entire title, sum up all the Word2vec's of the words present in the title and compute the average Word2vec by dividing the entire summed Word2vec with the total number of words present in the title. Similarly, compute the average Word2vec's for the queries, questions, and narratives in the same way giving rise to Average W2V model.

#### *4.3.1. Cosine Similarity for Average Word2vec*

Cosine similarity measures the angular distance between the two Average Word2vec's. The higher the cosine similarity, the more similar the title is. So, titles will be sorted for a particular query based on the cosine similarities in descending order. This will help us to know the titles that are closely relevant to the query. Upon predicting titles for a given set of queries, it resulted in 78% accuracy, which is much better than finding the similarities with the help of Bag-of-words. Likewise, repurposed input queries by taking the combinations of queries, questions, and narratives.

Heatmap of Word2vec model is given in Fig.6.

### *4.4. Average-BERT Based Model*

BERT is a model that has achieved a number of records for how well it can perform language-based tasks. BERT is basically a trained Transformer Encoder stack. The BERT Base model used for this covid dataset has 12 encoder layers. These have larger feedforward-networks with 768 hidden units and 12 attention heads. BERT receives a series of words as input, which continue to flow up the stack [9]. Each encoder performs self-attention, transfers the results through a feed-forward network, and then passes the information on to the next encoder. After reaching the last encoder the Bert base model will release the embeddings for each word present in the sentence based on the context. The output will be 768 dimensional word embedding for each word. So for every sentence the number of word tensors depends upon the number of words present in that sentence. But In-order to compare two sentences, the embedding for the entire sentence is required. For computing this, all the word tensors of each sentence are averaged to get 768 dimensional sentence embedding, which is unique in terms of dimension for every sentence in the dataset.

#### *4.4.1. Cosine Similarity for Average-Bert Based Vectors*

Cosine similarity is popular for finding better similarities of textual data. Here, the Cosine similarity is used to measure the angular distance between two Average BERT vectors. The higher the cosine similarity, the more similar the title is. So, titles will be sorted for a particular query based on the cosine similarities in descending order. This will help us to know the titles that are closely relevant to the query. As this model cares about both semantic and sequential information present in the textual data, the generated BERT base embeddings are more informative, which lead to better similarity i.e., high cosine similarity between query and question's Average BERT Base embeddings. It performed more precisely in finding relevant titles for the given queries when compared to Word to Vec, Bag of Words and Tf-Idf weighted Word2Vec.

The cosine similarity scores of question number 36 and query number 36 in topics dataset are given in Fig. 9.

### *4.5. TF-IDF Weighted Word2vec*

TF-IDF stands for Term Frequency and Inverse Document Frequency which is an NLP technique used to create vector representation for the given textual data. Firstly we will be calculating the TF values that gives us the information about how frequent the word Wj in the given review text Ti. Then IDF values are computed which gives the information about how less frequent or how rare a word occurs in Document Corpus. For Instance, If the word repetition is high in the review text, Term Frequency(TF) value increases. while, in case of Inverse Document frequency, IDF value is high if $W_j$ occurrence is rare in given document corpus. Both TF and IDF values are multiplied to get TF-IDF values. If TF and IDF values are high then TF-IDF will also be high. TF-IDF gives larger values for less frequent words in the document corpus. The TF-IDF values are computed by using the equation.2. After computing TF-IDF values for each word they are stored in the vector by considering words as feature names in the given review text.

$$TF - IDF = TF(W_j, T_i) * IDF(W_j, D)$$

$$TF(W_j, T_i) = \frac{\#\ times\ W_j\ occurs\ in\ T_i}{\#\ words\ in\ T_i}, \quad IDF(W_j, D) = log \frac{\#\ documents\ in\ D}{\#\ documents\ in\ D\ that\ Contain\ the\ word\ W_j} \quad (2)$$

After obtaining TF-IDF values for each word within every document, Word2Vecs are computed for each word in the text using the procedure described in Section.4.3. For each title and topic in the Data Corpus, two vectors are generated: the TF-IDF vector and the Word2Vec vector. After getting these two vectors, the TF-IDF values corresponding to each word in the text are considered as weights, and these weights are multiplied with each Word2Vec corresponding to each word in the relevant title or topic. After multiplying TF-IDF weights to Word2Vec vectors, add up all the TF-IDF weighted Word2Vec's of the words corresponding to the title, and divide the total TF-IDF weighted Word2Vec with the total number of words in the title to get the TF-IDF weighted Word2Vec for the entire title. Calculate the TF-IDF weighted Word2Vecs for the queries, questions, and narratives in the same way, As a result, a TF-IDF weighted Word2Vec model is created.

*4.5.1. Cosine Similarity for TF-IDF Weighted Word2Vec*

Cosine similarity measures the angular distance between the two TF-IDF weighted Word2Vec vectors. The higher the cosine similarity, the more similar the title is. So, titles will be sorted for a particular query based on the cosine similarities in descending order. This will help us to know the titles that are closely relevant to the query. Upon predicting titles for a given set of queries, it resulted in 79% accuracy, which is higher than finding the similarities with the help of Bag-of-words and Word2Vec. Furthermore, This model has given performance near to Average Bert Base model. Even though Bert-Base model has the highest performance in terms of accuracy, the TF-IDF weighted Word2Vec model has given the performance mostly near to Bert-Base model with lesser computational time. The comparison of both Bert-Base model and TF-IDF weighted Word2Vec model are given in the Table. 5.

*4.6. Performance Evaluation Measures*

*4.6.1. Accuracy*

Accuracy is used to evaluate the performance of the models used in this work. Accuracy is defined as

$$Accuracy = \frac{\#Correctly\ classified\ instances}{\#Instances\ in\ test\ dataset} \quad (3)$$

*4.6.2. Confusion Matrix*

Confusion matrix is used to measure the performance of models, where it computes True positive rate, False positive rate, True negative rate and False negative rate. These rates aid in determining where the model is sensibly perplexing. The positive and negative rates provided in the confusion matrix will aid in the computation of relevant metrics such as accuracy, precision, recall, and F1-Score, notably.

## 5. Results

In the competition of Trec-Covid, they provide the relevant judgments for some of the combinations of topics and titles. The given judgments are containing three labels:0,1 and 2 where 0 corresponds to non-relevant, 1 associates to partially-relevant and 2 is used to denote pure-relevant judgement. In this work, we converted this multi class problem to binary class problem by mapping non-relevant judgement to 0 and the other two to 1. We extract top 20 similar titles for every query based on the cosine similarity in bag-of-words, Word2vec, Average Bert-Base model, and Tf-Idf weighted W2V compared with the relevant judgements.

Here in this Research, heatmaps are used to visually represent the common words present in the title and query pairs. In case of bag of words the frequency of every word present in the title or query is represented which is shown in Fig.5. whereas in case of Word2vec, the cosine similarity between each word present in title is compared with every phrase present in respective query and is visually shown in Fig.6.

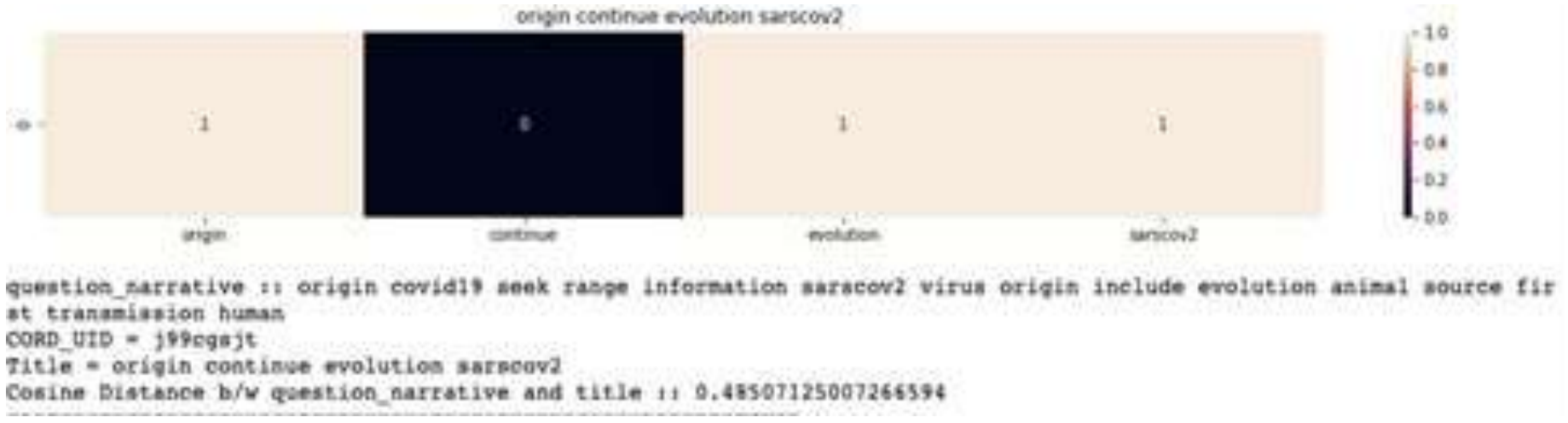

Figure 5: HeatMap of Bag-of-Words

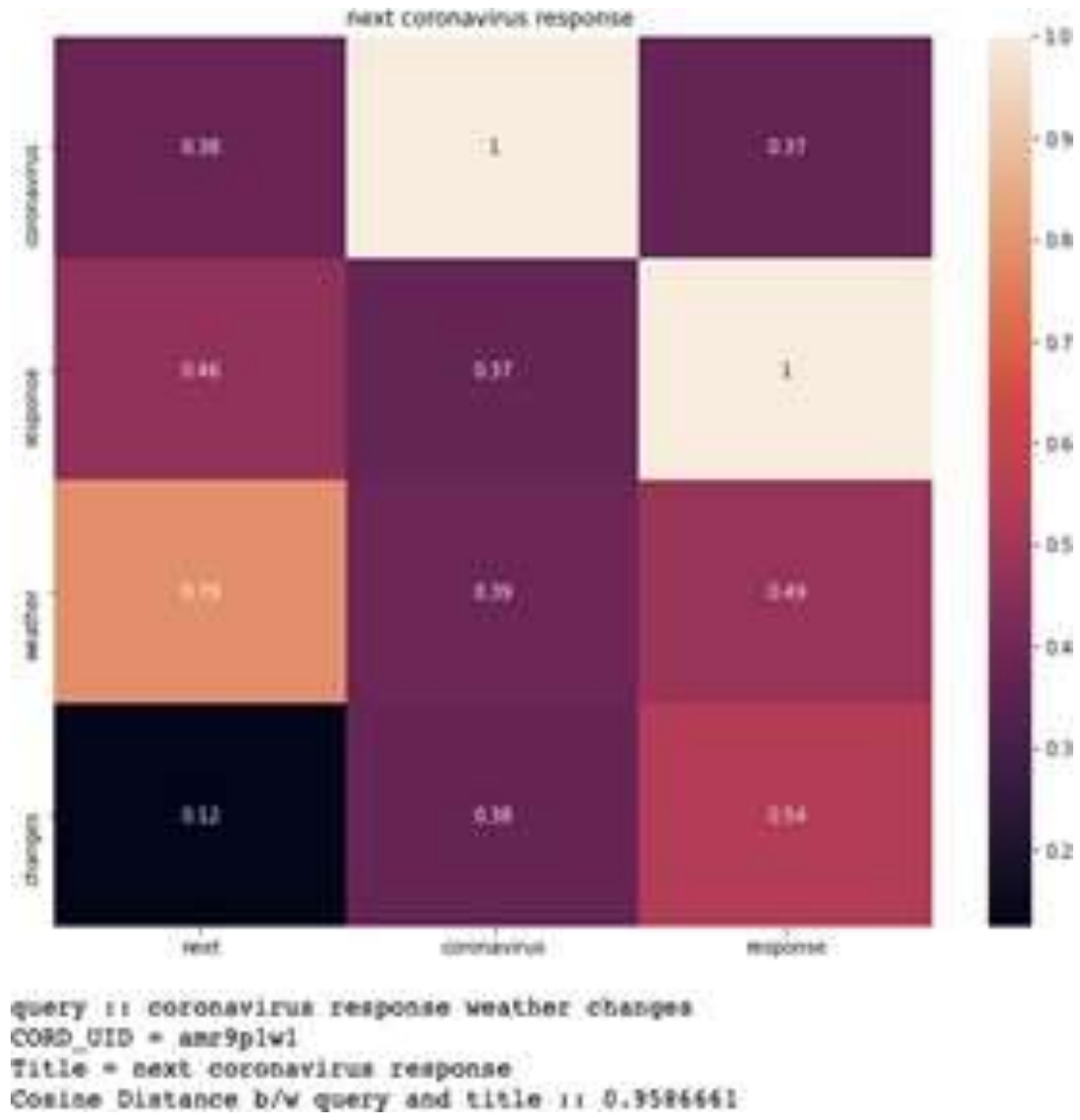

Figure 6: HeatMap of Word2Vec

We have taken threshold of cosine similarity scores by computing the mean using maximum cosine similarity value and minimum cosine similarity value, and present the accuracy of bag-of-words at threshold of 0.4 in Table.5.

Table 5: Accuracy of Bag of Words model

| TOPICS WITH TITLE | BOW ACCURACY WHEN THRESHOLD = 0.4 | BOW LEMMATIZED ACCURACY WHEN THRESHOLD = 0.4 |
|---|---|---|
| Query Title | 46% | 48% |
| Question Title | 24% | 29% |
| Narrative Title | 12% | 15% |
| Query Question Title | 36% | 41% |
| Query Narrative Title | 19% | 22% |
| Question Narrative Title | 16% | 20% |
| Query Question Narrative Title | 23% | 28% |

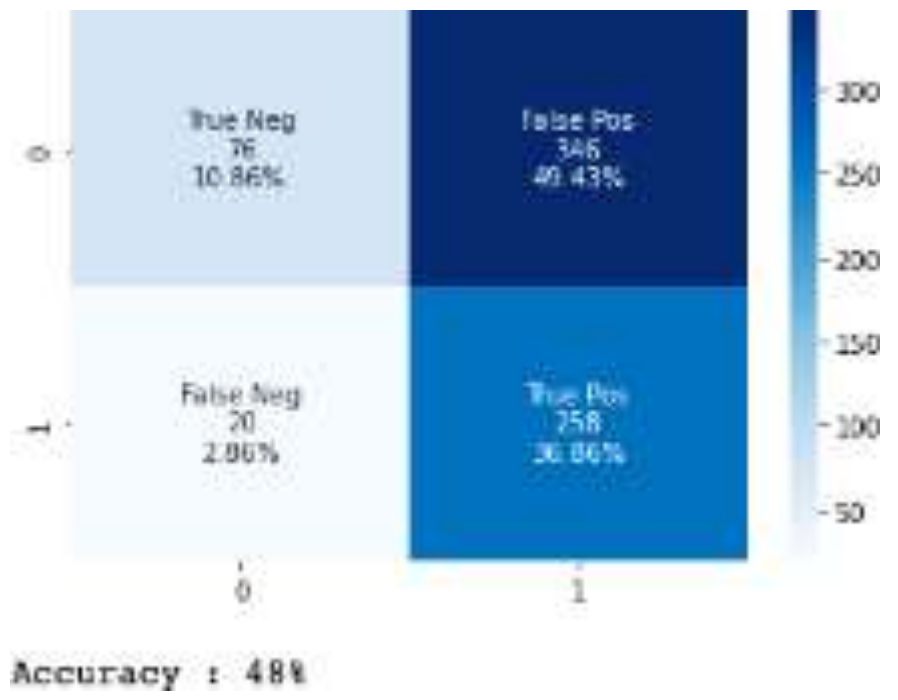

Figure 7: Confusion Matrix of Lemmatized Bag-of-Words model

The common observation in Bag-of-Words model is the accuracy is slightly improved by applying lemmatization to each word present in the given textual data. The highest accuracy of 48% is obtained when lemmatized Query title is used over question title and narrative title and the confusion matrix for query title is represented in Fig.7. Lemmatization improved the accuracy of frequency based bag of words model.

Accuracy of Word2vec for threshold of 0.45 is shown in Table.6.

Table 6: Accuracy of Word2Vec model

| TOPICS WITH TITLE | WORD2VEC ACCURACY | WORD2VEC LEMMATIZED ACCURACY |
|---|---|---|
| Query Title | 78% | 70% |
| Question Title | 74% | 70% |
| Narrative Title | 78% | 70% |
| Query Question Title | 77% | 71% |
| Query Narrative Title | 74% | 70% |
| Question Narrative Title | 77% | 67% |
| Query Question Narrative Title | 74% | 72% |

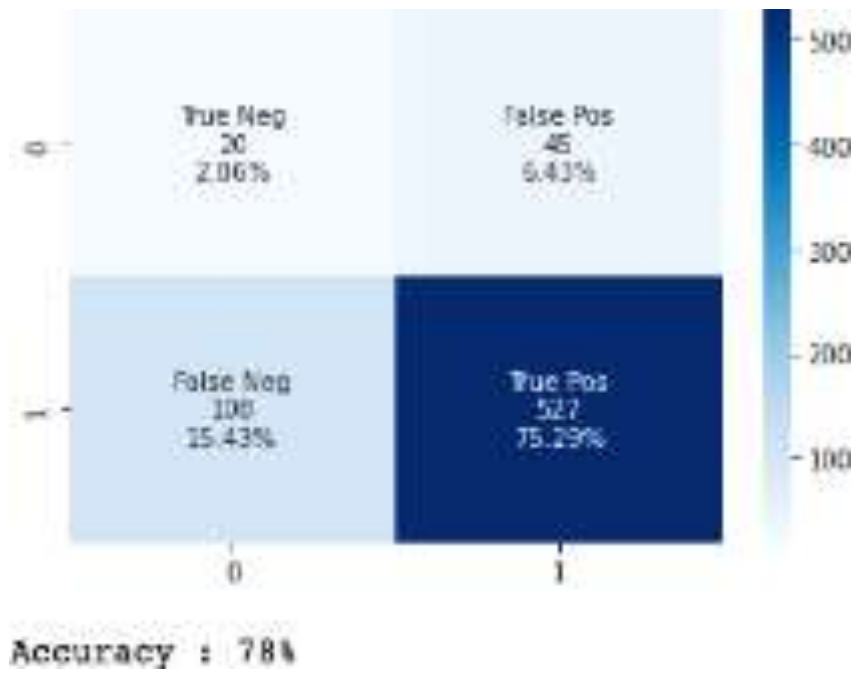

Figure 8: Confusion Matrix of Word2Vec

In Word2vec model, the accuracy is ranging from 74% - 78%. The highest accuracy is obtained for Narrative Title and Query Title and the confusion matrix for the narrative title is represented in Fig.8. But when lemmatization is applied to the text before giving the data to the Word2Vec model, the accuracy is lowering gradually. Therefore it is evident that Lemmatization is working better for frequency based models but not for the non frequency based models like Word2Vec.

| topic-id | cord-id | question | title | cosine_similarity |
|---|---|---|---|---|
| 36 | 3gpjsiak | protein structure sarscov2 spike | prefusion structure human coronavirus spike pr... | 0.894669 |
| 36 | 5oisrm5s | protein structure sarscov2 spike | identification common deletion spike protein s... | 0.894217 |
| 36 | cz3jq345 | protein structure sarscov2 spike | crystal structure orf9b lipid binding protein ... | 0.892966 |
| 36 | yvbsvalz | protein structure sarscov2 spike | model ace2 structure function sarscov receptor | 0.892733 |
| 36 | pm8usc55 | protein structure sarscov2 spike | insert sequence sarscov2 enhances spike protei... | 0.892675 |
| 36 | fsv5cuxe | protein structure sarscov2 spike | crystal structure s1 subunit nterminal domain ... | 0.892000 |
| 36 | 48ay8yl3 | protein structure sarscov2 spike | spike protein sarscov target vaccine therapeut... | 0.891300 |

| topic-id | cord-id | query | title | cosine_similarity |
|---|---|---|---|---|
| 36 | jqof906c | sarscov2 spike structure | expression sars spike gene shizomycete pombe | 0.863842 |
| 36 | 5ftpxlfe | sarscov2 spike structure | engineered stable miniprotein plug sarscov2 sp... | 0.858698 |
| 36 | sqz8yc7b | sarscov2 spike structure | neutralization sarscov2 destruction prefusion ... | 0.858262 |
| 36 | foolgmqj | sarscov2 spike structure | sars mers evidence speculation | 0.855996 |
| 36 | i9r77o70 | sarscov2 spike structure | characterization sarscov2 spike early prefusio... | 0.855639 |
| 36 | der39nfy | sarscov2 spike structure | silico map sarscov2 rna structurome | 0.855016 |
| 36 | u1e99bov | sarscov2 spike structure | identification application selfbinding zipper... | 0.854471 |
| 36 | 48ay8yl3 | sarscov2 spike structure | spike protein sarscov target vaccine therapeut... | 0.852883 |

Figure 9: Results of Average Bert-Base Model

Table 7: Accuracy of Average Bert-Base model

| TOPICS WITH TITLE | AVG BERT BASE ACCURACY |
|---|---|
| Query Title | 73% |
| Question Title | 90% |
| Narrative Title | 87% |
| Query Question Title | 89% |
| Query Narrative Title | 86% |
| Question Narrative Title | 88% |
| Query Question Narrative Title | 88% |

As seen in Fig.9 we have observed that Average Bert-Base Model given meaningful predictions for posed queries and questions and also resulted in higher accuracy than all other models as shown in Table .7. But this model takes more computational time for generating vectors from textual data, whereas TF-IDF weighted Word2Vec required much lesser computational time and it produced near results to Average Bert-Base.

TF-IDF Weighted Word2Vec Model.

Table 8: Accuracy of TF-IDF Weighted Word2Vec Model

| TOPICS WITH TITLE | TF-IDF Weighted Word2Vec ACCURACY |
|---|---|
| Query Title | 79% |
| Question Title | 87% |
| Narrative Title | 70% |
| Query Question Title | 85% |
| Query Narrative Title | 84% |
| Question Narrative Title | 84% |
| Query Question Narrative Title | 85% |

A Hybrid model which is a Combination of TF-IDF and Word2Vec models in-terms of vector fabrication helped in adding additional information to Word2Vec vectors by multiplying with TF-IDF weights. By adding the additional information to Word2Vec vectors for creating TF-IDF weighted Word2Vec resulted in better accuracy with less computational time when compared to Bag of words and Average Bert-Base model. This model achieved highest accuracy when posed Questions are compared with titles present in the document corpus which led to better recommendation of titles for the posed query.

Results of all NLP models used in this work are given in Table.

| TOPICS WITH TITLE | BoW | BoW + Lemmatization | W2V | W2V + Lemmatization | Bert | Tf-idf + W2V |
|---|---|---|---|---|---|---|
| Query Title | 46% | 48% | 78% | 70% | 73% | 79% |
| Question Title | 24% | 29% | 74% | 70% | 90% | 87% |
| Narrative Title | 12% | 15% | 78% | 70% | 87% | 70% |
| Query Question Title | 36% | 41% | 77% | 71% | 89% | 85% |
| Query Narrative Title | 16% | 22% | 74% | 70% | 86% | 84% |
| Question Narrative Title | 19% | 20% | 77% | 67% | 88% | 84% |
| Query Question Narrative Title | 23% | 28% | 74% | 72% | 88% | 85% |

## 6. Conclusion and future work

When comparing the accuracy of the bag of words model which is a frequency-based approach with the Word2vec model which understands the semantics of the words led us to the conclusion that the Word2vec model gave better predictions than the bag of words model. This happened due to the fact that the bag of words uses a frequency-based approach which only considers the higher occurring frequencies of the words instead of understanding the meaning of the query provided, which is exactly done in the Word2vec hence giving better results. In order to get more accurate results the meaning of the words in a sentence is not sufficient but the sequence information is also required. So the Bert base model having a stack of 12 encoders helps to fabricate Bert vectors with both semantic meaning and sequence information such that it leads to better accurate predictions more than the frequency based techniques and Average-Word2Vec. Even Average Bert-Base model has given higher performance than other two models, it takes more computational time when compared to other models which leads to more waiting time for getting document title recommendations for a posed topic which is not encouraged in real time applications. So, a Hybrid model is introduced which is a combination of TF-IDF and Word2Vec, this hybrid model gives the performance near to Average Bert-Base model and also computationally faster interms of recommending titles which is more useful. Our Future goal is to construct new hybrid models to achieve better and more accurate predictions.

## References

[1] Connor T Heaton and Prasenjit Mitra. “Repurposing TREC-COVID Annotations to Answer the Key Questions of CORD-19”. In: *arXiv preprint arXiv:2008.12353* (2020).

[2] Kaggle website. “Kaggle competetion.” In: *tps: // www. kaggle. com/ c/ trec-covid-information-retrieval* . 2020.

[3] Prakash M Nadkarni, Lucila Ohno-Machado, and Wendy W Chapman. “Natural language processing: an introduction”. In: *Journal of the American Medical Informatics Association* 18.5 (2011), pp. 544–551.

[4] Pedro Domingos. “A few useful things to know about Machine Learning”. In: *Communications of the ACM* 55.10 (2012), pp. 78–87.

[5] Ronan Collobert et al. “Natural language processing (almost) from scratch”. In: *Journal of machine learning research* 12.ARTICLE (2011), pp. 2493–2537.

[6] Long Ma and Yanqing Zhang. “Using Word2Vec to process big text data”. In: *2015 IEEE International Conference on Big Data (Big Data)*. IEEE. 2015, pp. 2895–2897.

[7] S Deepu, Pethuru Raj, and S Rajaraajeswari. “A Framework for Text Analytics using the Bag of Words (BoW) Model for Prediction”. In: *Proceedings of the 1st International Conference on Innovations in Computing & Networking (ICICN16), Bangalore, India*. 2016, pp. 12–13.

[8] Yin Zhang, Rong Jin, and Zhi-Hua Zhou. “Understanding bag-of-words model: a statistical framework”. In: *International Journal of Machine Learning and Cybernetics* 1.1 (2010), pp. 43–52.

[9] Jay Alammar. *The illustrated Bert, Elmo, and co.(how nlp cracked transfer learning)(2018)*. 2018.

[10] Komal Maher and Madhuri S Joshi. “Effectiveness of different similarity measures for text classification and clustering”. In: *International Journal of Computer Science and Information Technologies* 7.4 (2016), pp. 1715–1720.

[11] Bjornar Larsen and Chinatsu Aone. “Fast and effective text mining using linear-time document clustering”. In: *Proceedings of the fifth ACM SIGKDD international conference on Knowledge discovery and data mining*. 1999, pp. 16–22.

[12] Ricardo Baeza-Yates and R Ribeiro-Neto. *Modern Information Retrieval. NY*. 1999.

[13] Faisal Rahutomo, Teruaki Kitasuka, and Masayoshi Aritsugi. “Semantic Cosine Similarity”. In: 2012.

[14] Anjali Ganesh Jivani et al. “A Comparative Study of Stemming Algorithms”. In: *Int. J. Comp. Tech. Appl* 2.6 (2011), pp. 1930–1938.

[15] Jay Alammar. “The Illustrated Word2vec”. In: *Visualizing Machine Learning One Concept at a Time Blog* (2019).